\documentclass[12pt]{article}
\usepackage{graphicx}
\begin{document}
\centerline{\bf Monte Carlo Simulation of Medical Resource Allocation}

\medskip

Ana Proykova$^1$ and Dietrich Stauffer$^{1,2}$

\medskip
\noindent
$^1$ Department of Atomic Physics, University of Sofia, Sofia-1126, Bulgaria.

\noindent
$^2$ Institute for Theoretical Physics, Cologne University, 
D-50923 K\"oln, Euroland.

\medskip
Abstract: Computer simulations prove that we should spend more money on red 
wine than on medication. We optimise the problem of how to distribute a
fixed amount of money between medication and food spendings, using the
Penna ageing model. 

\medskip
How to spend your money is a difficult question and most easily solved by 
giving it, with all its worries, to us. As long as this is not yet done, one
has to balance at old age the need for medication and other medical support
against the need to buy healthy food and other goods needed for survival.
We simulate this problem here with the Penna ageing model \cite{books}, an
implementation of the mutation accumulation theory.

\begin{figure}[hbt]
\begin{center}
\includegraphics[angle=-90,scale=0.5]{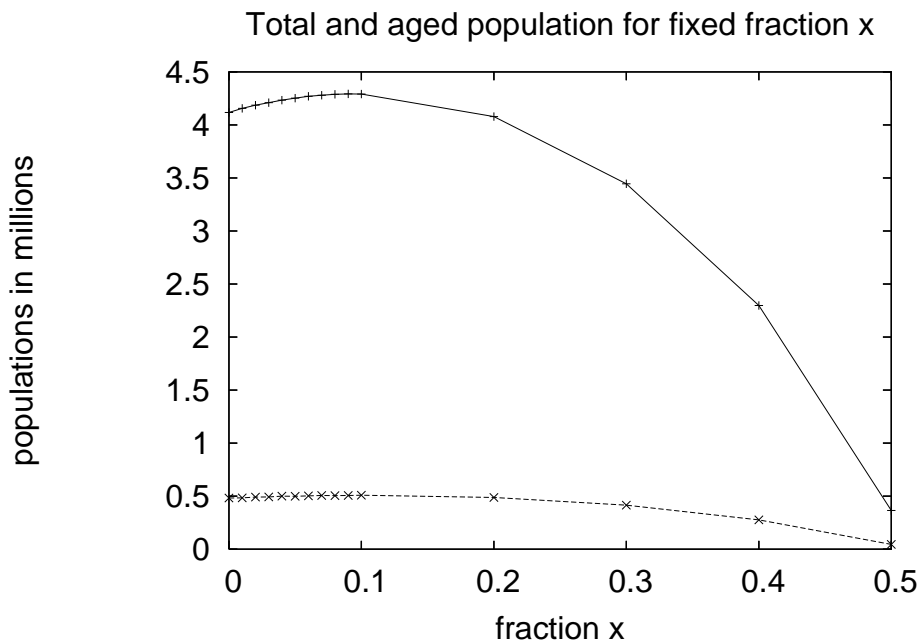}
\end{center}
\caption{ Variation of population with $x$ when the fraction $x$ of resources
is fixed for all. Average over the last 5000 of 10,000 time steps. The upper 
curve shows the total population, the lower one that with age above six.
}
\end{figure}
A fraction $H$ of individuals is healthy and does not need medication for 
survival. We assume the fraction $1-H$ of sick individuals to be their age $a$,
divided by 32, since 32 (in arbitrary time units, depending on the considered
species) is the maximum age: $a = 0,1,2,\dots,32$. (Our Penna model uses 
32 bits per bit-string.)
$$H = 1-a/32 \quad . \eqno (1) $$
The survival probability $p$ from one time step to the next is assumed to
be proportional to $(H+x)(1-x)$ and normalised to have a maximum of unity:
$$ p = 4(H+x)(1-x)/(1+H)^2 \quad . \eqno (2)$$
This maximum is reached for a fixed age at 
$$ x = (1-H)/2 \quad . \eqno(3) $$
Here $x$ is the fraction of money spent on medication; for $x=1$ nothing
is left and survival is impossible: $p(x=1) = 0$. 

A realistic population consists of individuals of many ages, and thus Eq.(3) 
is only a rough guide. We use the standard asexual Penna program \cite{books}
and add to it as another reason for death the above probability $1-p$ without
changing any of the other parameters. In a stationary population the maximum
age is then reached at 14; thus one time step corresponds to roughly 8 years
for humans.

Fig.1 shows the total population, as well as the population of individuals
above an age of six (roughly the oldest tenth of the population). We see 
a maximum at an optimal $x$ near 0.09, since the average $H$ is quite large.
This way of optimisation is appropriate for thinking people making decisions.

Mother Nature instead optimises by trial and error, using Darwinian survival 
of the fittest. We simulate this alternative by giving each individual its
own fraction $x$, which is given on to the offspring apart from small 
mutations which change $x$ at birth by a random number between --0.01 and
+0.01. Then Fig.2 shows how the average $x$ evolves from its initial value
1/2 ($0 \le x \le 1$ randomly at start) to about the same optimal value
0.09 seen already from Fig.1.

Feedback is introduced by replacing in Eq.(1) the {\it a priori} fraction
age/32 of sick people by $M/T$ where $M$ counts the number of active mutations
(hereditary diseases) and $T=3$ their threshold: Three active diseases kill.
Then after some time the average $x$ may increase again. Fig.3, with a 
bimodal distribution for $x$ at intermediate times. 

At a time when old-age pensions no longer grow as much as they did in the past,
simulations like these may guide us how to spend them best.

\begin{figure}[hbt]
\begin{center}
\includegraphics[angle=-90,scale=0.5]{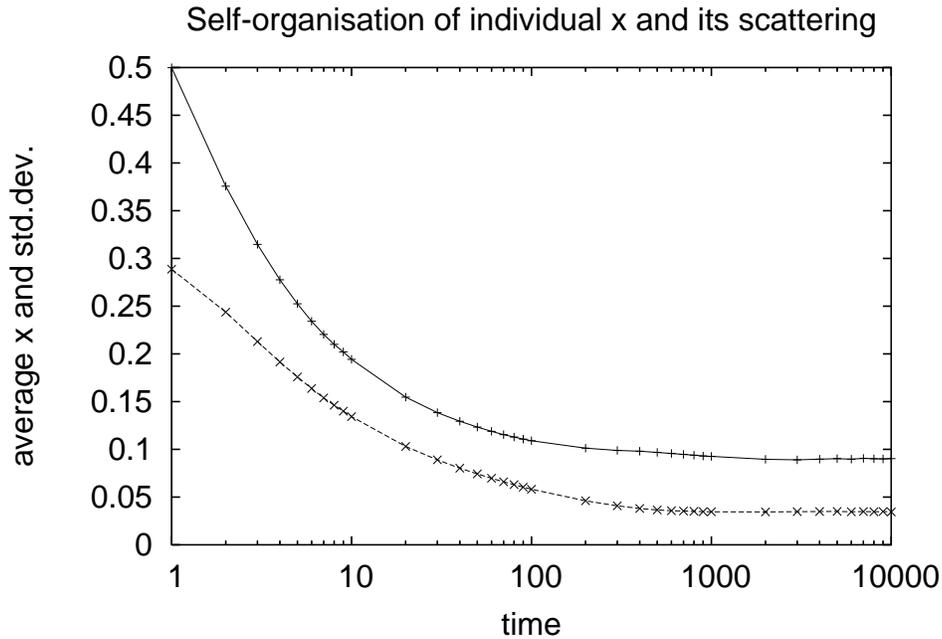}
\end{center}
\caption{Self-organisation of individual $x$ values towards the optimum.
The upper data show the average $<x>$, the lower its standard deviation 
$(<x^2> - <x>^2)^{1/2}$. The final population was 4.3 million similar to Fig.1.
}
\end{figure}

\begin{figure}[hbt]
\begin{center}
\includegraphics[angle=-90,scale=0.45]{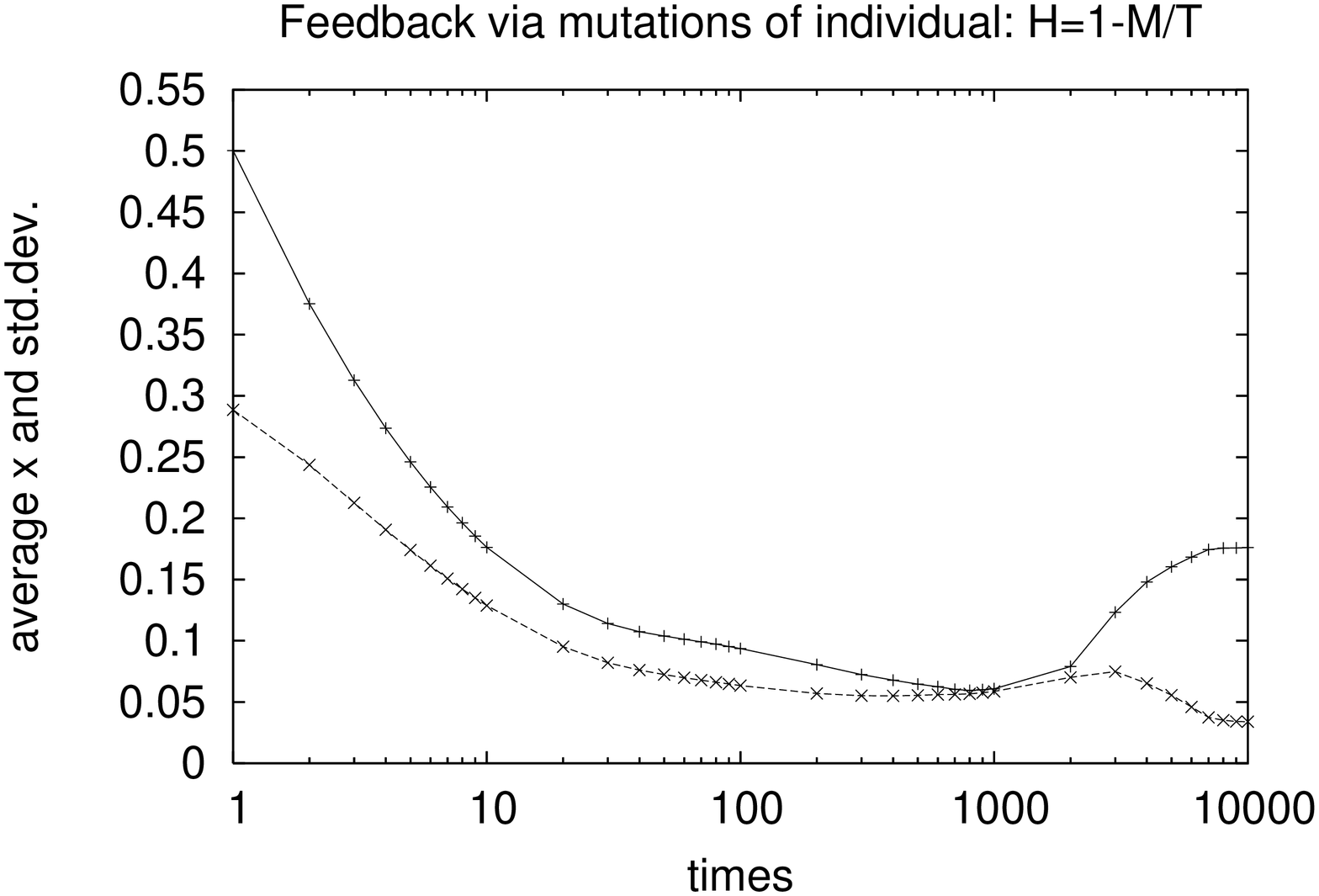}
\includegraphics[angle=-90,scale=0.45]{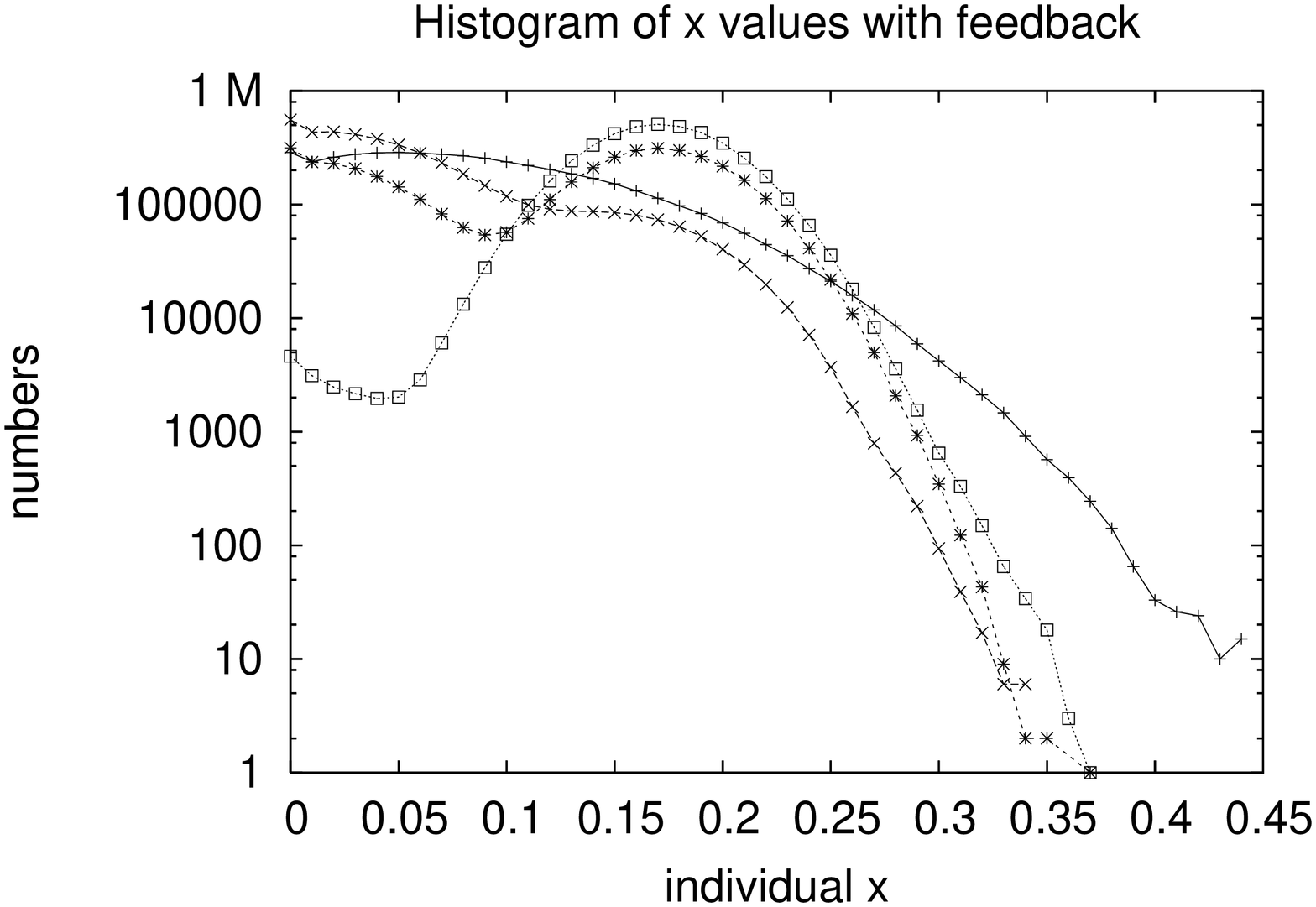}
\end{center}
\caption{Top: As Fig.2 but with feedback $H = 1-M/T$; only minor changes
for $10,000 \le t \le 20,000$ (not shown). Bottom: Distribution of $x$ 
values in the same simulation, after $t = 100$ (+), 500 (x,left), 3000
(*,middle), and 5000 (squares, right) iterations. 
}
\end{figure}

\end{document}